\documentclass[12pt]{article}
\usepackage{cite}
\usepackage[utf8]{inputenc}
\usepackage{epsf,amsmath,epsfig,graphicx,dcolumn}
\usepackage{scalefnt,ulem,pstricks}
\usepackage{rotating}
\usepackage{epstopdf}
\usepackage{color}
\usepackage{subfigure}
\usepackage{tikz-feynman}
\newcommand\as{\alpha_{\mathrm{S}}}
\newcommand\gs{g_{\mathrm{S}}} 
\newcommand\f[2]{\frac{#1}{#2}}
\newcommand{\calQ}{\mathcal{Q}}
\newcommand{\nn}{\nonumber}
\newcommand{\mh}{m_H}

\newcommand{\muR}{\mu_{R}}

\definecolor{mygreen}{rgb}{0.13, 0.55, 0.13}
\definecolor{myviolet}{rgb}{0.6, 0.4, 0.8}

\begin{document}
\begin{titlepage}
  \renewcommand{\thefootnote}{\fnsymbol{footnote}}
  \begin{flushright}
    ZU-TH 21/18 \\
    PSI-PR-18-06
  \end{flushright}
  \vskip 1cm
\begin{center}
{\Large \bf Higgs boson production at large transverse momentum\\[0.2cm]
within SMEFT: analytical results}
\end{center}

\par \vspace{2mm}
\begin{center}
{\bf Massimiliano Grazzini$^{(a)}$},
{\bf Agnieszka Ilnicka$^{(a,b,c)}$},
{\bf Michael Spira$^{(c)}$} 

\vspace{5mm}

$^{(a)}$ Physik-Institut, Universit\"at Z\"urich, 
CH-8057 Z\"urich, Switzerland 

$^{(b)}$ Physics Department, ETH Z\"urich, 
CH-8093 Z\"urich, Switzerland 

$^{(c)}$ Paul Scherrer Institute, CH-5232 Villigen PSI, Switzwerland

\end{center}

\par \vspace{2mm}
\begin{center} {\large \bf Abstract} \end{center}
\begin{quote}
\pretolerance 10000

We consider Higgs boson production through gluon fusion at large transverse momentum in hadronic collisions.
We present the analytic expressions of the relevant one-loop QCD amplitudes including the effects of the complete set of dimension-six operators.
The latter correspond to modifications of the top and bottom Yukawa couplings, to an effective point-like Higgs coupling to gluons and
to the chromomagnetic operator of the top quark. The quantitative impact of the chromomagnetic operator is also studied.
Our results confirm previous findings that the effect of the
chromomagnetic operator at high $p_T$ can be large and should not be neglected.
\end{quote}

\vspace*{\fill}
\begin{flushleft}
June 2018
\end{flushleft}
\end{titlepage}

\section{Introduction}
%        ============
After the discovery of the scalar resonance of mass 125 GeV
\cite{ATLASdisc,CMSdisc} the measurement of its properties is one of the
main activities of the LHC program. The Run I measurements
\cite{Khachatryan:2016vau} showed that the new resonance is compatible with the
Standard Model Higgs boson. There is, however, still the possibility that more precise
measurements will uncover small deviations from the Standard Model (SM) predictions. These might be the long
lasting legacy of the LHC, which will encompass the searches for New
Physics. The need of a consistent framework to capture small deviations from the SM is reflected
in the wide discussions in Refs.~\cite{Heinemeyer:2013tqa,YR4,HTBook}.
The Standard Model Effective Field Theory (SMEFT) is a promising and
theory motivated approach, in which the deviations from the SM are
parametrised with higher-dimension operators, in the first
approximation dimension six \cite{dim61,dim62}. 

Next to the inclusive quantities, differential Higgs observables were measured in Run I \cite{atlas1,atlas2,Aad:2015lha,Aad:2016lvc,CMSpt, Khachatryan:2015yvw,Khachatryan:2016vnn}
and with a partial data set of Run II \cite{Aaboud:2017oem,Aaboud:2018xdt,Sirunyan:2017exp}, although still with relatively large uncertainties.
With the
increasing amount of collected data, the statistical accuracy will
improve, thereby allowing us to put stringent constraints on the SMEFT parameters.
One of the observables which is able to shed light on the structure of the Higgs sector is the transverse momentum spectrum ($p_T$) of the Higgs particle.
For example, a measurement of the $p_T$ spectrum could give insight on the nature of the Higgs boson coupling to gluons (see e.g. Refs.~\cite{Banfi:2013yoa,Azatov:2016xik}).

Dedicated calculations and tools are needed to enable the experimental
analyses to set bounds on the SMEFT operators. Approximate
results for the total gluon fusion Higgs production cross section
including modified top and bottom Yukawa couplings and an additional
direct $Hgg$ interaction have been obtained at NNLO in QCD perturbation theory
in Ref.~\cite{Brooijmans:2016vro} and at N$^3$LO in
Refs.~\cite{Harlander:2016hcx,Anastasiou:2016hlm}. As far as gluon fusion is concerned,
the inclusion of dimension-six and dimension-eight operators in the
Higgs $p_T$-spectrum also has been considered in
Refs.~\cite{ptdim61,ptdim62,ptdim63} and \cite{ptdim81,ptdim82},
respectively. Strategies for extracting information on the Higgs-gluon
couplings from the measurements were studied in Ref.~\cite{ptdim63}, and
the study the low-$p_T$ range therein was made possible by using Monte Carlo Parton Shower.
Also in Ref.~\cite{Azatov:2016xik} the prospects of the determination of
the Wilson coefficients in the high-luminosity LHC and future colliders were
considered. The mentioned studies usually omitted the effects of the
chromomagnetic operator, but a dedicated work analysed its effect on the
LO Higgs production \cite{Choudhury:2012np}. This was followed by a LO
study \cite{Degrande:2012gr} on the interplay of the SMEFT operators
entering top-induced Higgs production channels, with the
chromomagnetic operator treated in the heavy-top limit (HTL). Recently,
the program of the SMEFT at NLO QCD was started by the
{\sc MadGraph5\_aMC@NLO} group \cite{Zhang:2016snc} and %in the context of Higgs physics
led to the calculation of $t\bar tH, tH$
\cite{Maltoni:2016yxb} and recently also of Higgs production through gluon fusion \cite{Deutschmann:2017qum}.

In this work we recall the results for the LO Higgs production via gluon
fusion and we extend our study \cite{ours} of the Higgs $p_T$ spectrum
to include the effects of the chromomagnetic operator. More precisely, we present the
analytic expressions of the relevant one-loop QCD amplitudes including the effects of the complete set of dimension-six operators and
we shortly illustrate the impact of the chromomagnetic operator on the high-$p_T$ tail of the spectrum.
Note that, due to the automated character of the calculations in the {\sc MadGraph5\_aMC@NLO} framework \cite{Maltoni:2016yxb,Deutschmann:2017qum},
they can be considered complementary
to the analytic calculations presented here. %A more extensive phenomenological study will be presented elsewhere.

The paper is organised as follows. In Sect.~1 we review the LO results and set up our notation. In Sect.~2 we present the analytical results for the SMEFT  one-loop QCD amplitudes
in all partonic channels, and we briefly discuss the impact of the chromomagnetic operator at high $p_T$. In Sect.~4 we draw our conclusions.

\section{Framework and LO results}
%        ========================
We consider the effective Lagrangian
\begin{equation}
  {\cal L}={\cal L}_{SM}+\sum_i \f{c_i}{\Lambda^2}{\cal O}_i
\end{equation}
where the SM is supplemented by the inclusion of a set of dimension-six
operators describing new physics effects at a scale $\Lambda$ well above
the EW scale. We focus on the following three operators
\begin{equation}
  \label{eq:operators}
{\cal O}_1 = |H|^2 \bar{q}_L H^c t_R + h.c.\quad\quad\quad  {\cal O}_2 = |H|^2 G^a_{\mu\nu}G^{a,\mu\nu}\quad\quad\quad {\cal O}_3 = \bar{Q}_L H \sigma^{\mu\nu}T^a t_R G_{\mu\nu}^a + h.c.\  .
\end{equation}
These operators, in the case of single Higgs production, may be rewritten as:
\begin{align}
  \label{eq:o1}
  \f{c_1}{\Lambda^2}\,{\cal O}_1 &\rightarrow c_1 \frac{m_t}{v} h \bar{t} t\,,\\
\label{eq:o2}  
\f{c_2}{\Lambda^2}\,{\cal O}_2 &\rightarrow c_2 \frac{\as}{\pi v} h  G^a_{\mu\nu}G^{a,\mu\nu}\,,\\
\label{eq:o3}
\f{c_3}{\Lambda^2}\,{\cal O}_3 &\rightarrow c_{3}\f{\gs m_t}{2v^3} (v+h)G_{\mu\nu}^a({\bar t}_L\sigma^{\mu\nu}T^a t_R+h.c),
\end{align}
where $\as$ is the QCD coupling ($\as=\gs^2/(4\pi)$), $m_t$ is the (pole) mass of the top quark,
$v$ is the expectation value of the Higgs field, $v=(\sqrt{2}G_F)^{-1/2}\sim 246$ GeV
and $\sigma^{\mu\nu}=\f{i}{2}\left[\gamma^\mu,\gamma^\nu\right]$.
The operator ${\cal O}_1$ is the Yukawa operator, and describes
modifications of the $t\bar tH$ coupling.  The operator ${\cal O}_2$
provides a contact interaction of the Higgs boson and gluons with the
same structure as in the heavy-top limit of the SM.  The operator ${\cal
O}_3$ is the chromomagnetic dipole moment operator, which modifies the
interactions between gluons and the top quark.  In our
convention, based on the SILH basis \cite{SILH0,SILH}, we express the
Wilson coefficients as factors in the canonically normalized Lagrangian. 

To set up our convention we reproduce the results for the LO inclusive
cross section for $gg\to H$ as exemplified in Refs.~\cite{ours,
our_Moriond, our_discrete}. After renormalizing the point-like
Higgs-gluon coupling $c_2$ in the ${\overline {\rm MS}}$ scheme the LO matrix element can be decomposed as
\begin{equation}
{\cal T}_{gg\to H}(p_1,p_2)=i\f{\as}{3\pi
v}\epsilon_{\mu}(p_1)\epsilon_{\nu}(p_2)
\left[p_1^\nu p_2^\mu-(p_1p_2)g^{\mu\nu}\right]F(\tau_H)\, ,
\end{equation}
where $p_1$, $p_2$ are the gluon momenta, $\epsilon(p_1)$, $\epsilon(p_2)$ their polarisations and $\tau_H=4m_t^2/\mh^2$, $\mh$ being the Higgs boson mass. The form factor $F(\tau)$ is defined as
\begin{equation}
F(\tau)=c_1 F_1(\tau)+c_2(\muR) F_2(\tau)+Re(c_3)\f{m_t^2}{v^2} F_{3}(\tau)\, ,
\end{equation}
with\footnote{Note that we changed our sign convention of $c_3$
compared to Ref.~\cite{ours}.}
\begin{align}
F_1(\tau)&=\f{3}{2}\tau \left[1+(1-\tau)f(\tau)\right]\,,\\
F_2(\tau)&=12\, ,\\
F_3(\tau)&=3\left(\tau f(\tau)+2g(\tau)-1-2\ln \f{\muR^2}{m_t^2}\right)\,
\end{align}
and the functions $f,g$ are defined in Appendix A. In the HTL the form
factors approach the simple expressions
\begin{align}
F_1(\tau)& \to 1\,,\\
F_2(\tau)& \to 12\, ,\\
F_3(\tau)& \to 6\left( 1-\ln \f{\muR^2}{m_t^2}\right) \, .
\end{align}

\section{Higgs plus jet production}
\begin{figure}[h]
\begin{center}
\includegraphics[width=0.5\textwidth]{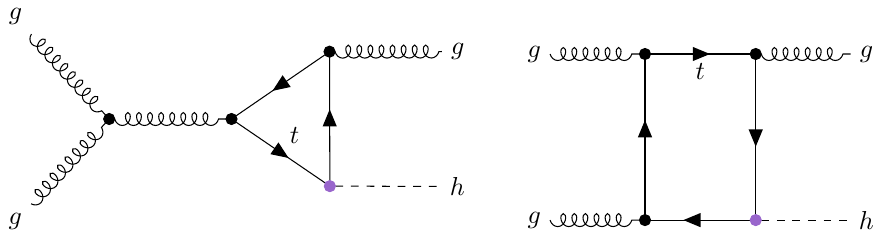}
\end{center}
\caption{\label{fig:SM} Generic diagrams originating from the
${\cal O}_1$ operator that also provide the SM contribution.}
\end{figure}

%        =========================
Higgs boson production in association with a jet is the LO contribution
to Higgs boson production at finite transverse momenta. This process is
mediated by $gg, gq$ and $q\bar q$ initial states. We start the
presentation of our results for the $gg$ channel,
\begin{equation}
  g(p_1)+g(p_2)\to g(p_3)+H(q)\nn
\end{equation}
and the Mandelstam variables are defined as
\begin{equation}
  s=2p_1\cdot p_2\quad\quad\quad t=-2p_1\cdot p_3\quad\quad\quad u=-2p_2\cdot p_3\quad\quad\quad {\rm with}\quad\quad\quad s+t+u=\mh^2\, .
  \end{equation}
The contributing generic SM diagrams are shown in Fig.~\ref{fig:SM}.

\begin{figure}[h]
\begin{center}
\includegraphics[width=0.5\textwidth]{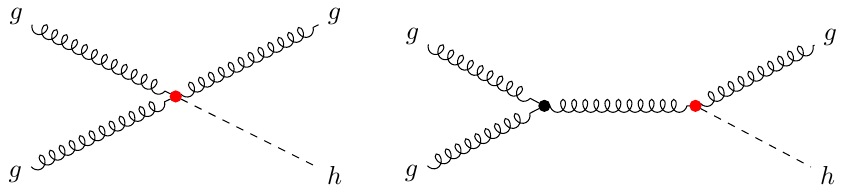}
\end{center}
\caption{\label{fig:SMeff} Generic diagrams originating from
the ${\cal O}_2$ operator. These correspond also to the effective SM
diagrams in the HTL.}
\end{figure}

\begin{figure}[h]
\begin{center}
\includegraphics[width=\textwidth]{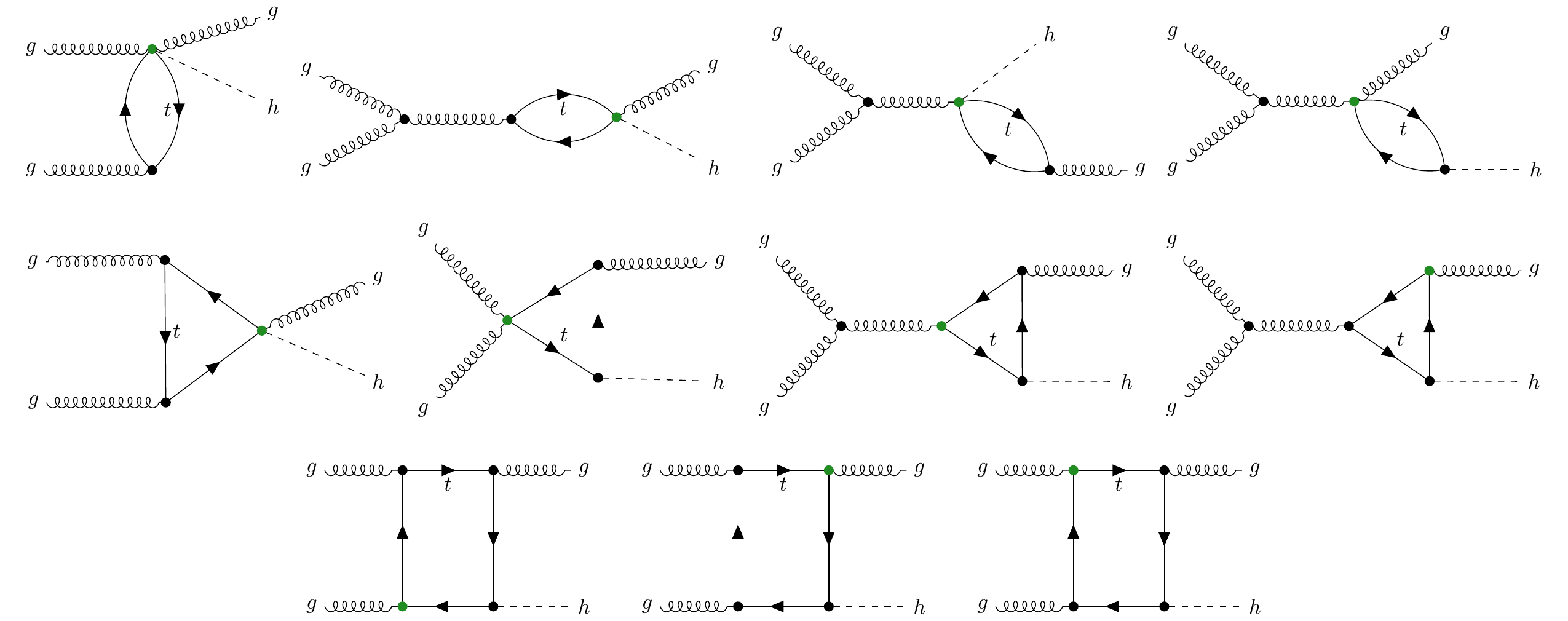}
\end{center}
\caption{\label{fig:chromo} Generic diagrams induced by the
chromomagnetic operator ${\cal O}_3$.}
\end{figure}

The contribution from the modified Yukawa coupling can be
straightforwardly obtained by rescaling the SM result.  The
effective Higgs-gluon coupling gives rise to the diagrams in
Fig.~\ref{fig:SMeff}.
When considering the insertion of the chromomagnetic operator we obtain
54 additional diagrams (see Fig.~\ref{fig:chromo}) out of which just 2
types are topologically equivalent to the SM ones.

The amplitude can be cast into the form
\begin{equation}
\mathcal{T}_{gg\to gH}(p_1,p_2,p_3) = \frac{ig_S^3 \mh^4}{12 \pi^2 v}
f_{abc} \mathcal{M}_{gg\to gH}^{\mu \nu \rho}  \epsilon_{\mu}(p_1)
\epsilon_{\nu}(p_2) \epsilon^*_{\rho}(p_3) \\
\end{equation}
where $\epsilon(p_i)$ are the gluon polarisation vectors.
%\aicom{I'm not sure if this technical part is needed, not just that $\mathcal{M}_{gg\to gH}^2 = C_1^2+ C_2^2+ ...$ and then definitions of $C_i$ }
The amplitude can be decomposed into four independent tensor structures \cite{Spira:1995es} 
\begin{align}
  \label{eq:deco}
  \begin{split}
\mathcal{M}_{gg\to gH}^{\mu \nu \rho}(p_1,p_2,p_3) & = F_1(p_1,p_2,p_3) \calQ^{\mu \nu \rho}_1 + F_2(p_1,p_2,p_3) \calQ^{\mu \nu \rho}_2 \\
& + F_3(p_1,p_2,p_3) \calQ^{\mu \nu \rho}_3 + F_4(p_1,p_2,p_3) \calQ^{\mu \nu \rho}_4
\end{split}
\end{align}
where
\begin{align}
\begin{split}
\calQ^{\mu \nu \rho}_1 & =  p_1^{\rho} p_2^{\mu} p_3^{\nu} - p_1^{\nu} p_2^{\rho} p_3^{\mu} + g^{\mu \nu} [ (p_1 \cdot p_3) p_2^{\rho} - (p_2 \cdot p_3) p_1^{\rho} ] \\
 & + g^{\mu \rho} [ (p_2 \cdot p_3) p_1^{\nu} - (p_1 \cdot p_2) p_3^{\nu} ] + g^{\nu \rho} [ (p_1 \cdot p_2) p_3^{\mu} - (p_1 \cdot p_3) p_2^{\mu} ]   \\
\calQ^{\mu \nu \rho}_2 & = [ (p_2 \cdot p_3) p_1^{\rho} - (p_1 \cdot p_3) p_2^{\rho} ] \frac{p_1^{\nu}p_2^{\mu} - (p_1 \cdot p_2) g^{\mu \nu}}{(p_1 \cdot p_2)} \\
\calQ^{\mu \nu \rho}_3 & = [ (p_2 \cdot p_3) p_1^{\nu} - (p_1 \cdot p_2) p_3^{\nu} ] \frac{p_1^{\rho}p_3^{\mu} - (p_1 \cdot p_3) g^{\mu \rho}}{(p_1 \cdot p_3)} \\
\calQ^{\mu \nu \rho}_4 & = [ (p_1 \cdot p_3) p_2^{\mu} - (p_1 \cdot p_2) p_3^{\mu} ] \frac{p_2^{\rho}p_3^{\nu} - (p_2 \cdot p_3) g^{\nu \rho}}{(p_2 \cdot p_3)}\, .
\end{split}
\end{align}
It is useful to note that based on the definitions of the tensor
structures the form factors have the following properties due to Bose
symmetry,
\begin{equation}
F_2(p_1,p_2,p_3) =  F_2(p_2,p_1,p_3) = - F_3(p_1,p_3,p_2) =  F_4(p_3,p_2,p_1) 
\end{equation}
and $F_1$ is totally symmetric. When squaring the amplitude, the
decomposition in Eq.~(\ref{eq:deco}) leads to mixed terms in the form
factors $F_i$.  If we rearrange the form factors as
\begin{align}
\begin{split}
C_1(s,t,u;m_t) &= \frac{stu}{2}~[2 F_1(p_1,p_2,p_3) + F_2(p_1,p_2,p_3)
- F_3(p_1,p_2,p_3) + F_4(p_1,p_2,p_3)]\\
C_2(s,t,u;m_t) &= \frac{stu}{2}~F_2(p_1,p_2,p_3)\\
C_3(s,t,u;m_t) &= \frac{stu}{2}~F_3(p_1,p_2,p_3)\\
C_4(s,t,u;m_t) &= \frac{stu}{2}~F_4(p_1,p_2,p_3)\, ,
\end{split}
\end{align}
where the $C_i$ develop the analogous Bose-symmetry properties as the
previous form factors
\begin{align}
\begin{split}\label{eq:Cperm}
C_3(s,t,u;m_t) &= -C_2(t,s,u;m_t)\\
C_4(s,t,u;m_t) &= C_2(u,t,s;m_t)
\end{split}
\end{align}
the amplitude squared can now be expressed as a sum of absolute squares
\begin{equation}
|\mathcal{T}_{gg\to gH}|^2 = \frac{32\alpha_S^3}{3 \pi v^2}
\frac{\mh^8}{s t u} (|C_1|^2+|C_2|^2+|C_3|^2+|C_4|^2).
\end{equation}
Here we present the results for $C_1(s,t,u;m_t)$ and $C_2(s,t,u;m_t)$
($C_3$ and $C_4$ can be obtained from $C_2$ following Eq.
(\ref{eq:Cperm})) in terms of the scalar integrals:
\begin{equation}
C_i^{(\alpha)} = \frac{1}{32}\tau_H^2\sum_{j=1}^{12} P^{(\alpha)}_{i,j} T_j, 
\end{equation}
where the $\alpha$ index runs over the operators ${\cal O}_\alpha$ in Eq.~(\ref{eq:operators}) and
\begin{align}
%\begin{array}{llll}
T_1 &= 1 & T_2 &= 2\left[1-g\left(\tau_s\right)\right] \\
T_3 &= 2\left[1-g\left(\tau_t\right)\right] & T_4 &= 2\left[1-g\left(\tau_u\right)\right] \\
T_5 &= 2[1-g(\tau_H)] & T_6 &= 2 f(\tau_s) \\
T_7 &= 2 f(\tau_t) & T_8 &= 2 f(\tau_u) \\
T_9 &= 2 f(\tau_H) & T_{10} &= J(\tilde{s},\tilde{t},\tilde{u}) \\
T_{11} &= J(\tilde{s},\tilde{u},\tilde{t}) & T_{12} &= J(\tilde{u},\tilde{s},\tilde{t})\, . \\
%\end{array}
\end{align}
In order to present our results in a compact form we have defined dimensionless quantities
$\tilde{x} = x/m_t^2$ and $\tau_x=4/\tilde{x}$ for $x=s,t,u$ and $\rho=\mh^2/m_t^2$. The scalar functions $f(x)$, $g(x)$ and
$J(x,y,z)$ are given in Appendix~\ref{sec:scalint}. The $P_{i,j}^{(1)}$
coefficients, corresponding to the SM contribution, read
\begin{equation}
\begin{array}{llll}
P_{1,1}^{(1)} &= 12 \rho & P_{1,2}^{(1)} &= 0 \\
P_{1,3}^{(1)} &= 0 & P_{1,4}^{(1)} &= 0 \\
P_{1,5}^{(1)} &= 0 & P_{1,6}^{(1)} &= 3 (4 - \rho) \\
P_{1,7}^{(1)} &= 3 (4 - \rho) & P_{1,8}^{(1)} &= 3 (4 - \rho) \\
P_{1,9}^{(1)} &= -9 (4 - \rho) & P_{1,10}^{(1)} &= \frac{3}{2} \tilde{s} \tilde{u} (4 - \rho) \\
P_{1,11}^{(1)} &= \frac{3}{2} \tilde{s} \tilde{t} (4 - \rho) & P_{1,12}^{(1)} &= \frac{3}{2} \tilde{t} \tilde{u} (4 - \rho) \\
\end{array}
\end{equation}
\begin{align}\label{eq:pct}
\begin{split}
P_{2,1}^{(1)} &= - 12 \tilde{s} \frac{\tilde{u}\tilde{t} - \tilde{s}^2}{(\tilde{s}+\tilde{u})(\tilde{s}+\tilde{t})} \\
P_{2,2}^{(1)} &= 0 \\
P_{2,3}^{(1)} &= \frac{12 \tilde{t} \tilde{u} \left(\tilde{u} + 2\tilde{s}\right)}{(\tilde{s}+\tilde{u})^2} \\
P_{2,4}^{(1)} &= \frac{12 \tilde{t} \tilde{u} \left(\tilde{t} + 2\tilde{s}\right)}{(\tilde{s}+\tilde{t})^2} \\
%P_{2,5}^{(1)} &= -\frac{12 \tilde{t}\tilde{u}}{(\tilde{s}+\tilde{t})^2 (\tilde{s}+\tilde{u})^2} (\tilde{u}^2\tilde{t} + 2 \tilde{u}^2 \tilde{s} + \tilde{u}\tilde{t}^2 + 4\tilde{u}\tilde{t}\tilde{s} + 5\tilde{u}\tilde{s}^2 + 2\tilde{t}^2\tilde{s} + 5 \tilde{t}\tilde{s}^2 + 4\tilde{s}^3)\\
P_{2,5}^{(1)} &= -\frac{12 \tilde{t}\tilde{u}}{(\tilde{s}+\tilde{t})^2
(\tilde{s}+\tilde{u})^2}\left[(3\tilde{s}^2+\tilde{t}\tilde{u}
+2\rho\tilde{s})(\rho-\tilde{s})+4\tilde{s}^3\right] \\
P_{2,6}^{(1)} &= -3 (\tilde{s} - 4)\\
%P_{2,7}^{(1)} &= -\frac{3 \left(4\tilde{u}^3 \tilde{t} + 8\tilde{u}^2\tilde{t}\tilde{s} - \tilde{u}^2\tilde{s}^2 + 4\tilde{u}^2\tilde{s} + 4\tilde{u}\tilde{t}\tilde{s}^2 + 8\tilde{u}\tilde{s}^2 + \tilde{s}^4 - 4\tilde{s}^3 \right)}{\tilde{s} (\tilde{s}+\tilde{u})^2}\\
P_{2,7}^{(1)} &= -\frac{3 \left(4\tilde{t}\tilde{u}(\tilde{s}+\tilde{u})^2
+\tilde{s}(4-\tilde{s})(\tilde{u}^2-\tilde{s}^2)+8\tilde{s}^2\tilde{u}\right)}
{\tilde{s}(\tilde{s}+\tilde{u})^2}\\
%P_{2,8}^{(1)} &= -\frac{3 \left(4\tilde{t}^3 \tilde{u} + 8\tilde{t}^2\tilde{u}\tilde{s} - \tilde{t}^2\tilde{s}^2 + 4\tilde{t}^2\tilde{s} + 4\tilde{u}\tilde{t}\tilde{s}^2 + 8\tilde{t}\tilde{s}^2 + \tilde{s}^4 - 4\tilde{s}^3\right)}{\tilde{s} (\tilde{s}+\tilde{t})^2}\\
P_{2,8}^{(1)} &= -\frac{3 \left(4\tilde{t}\tilde{u}(\tilde{s}+\tilde{t})^2
+\tilde{s}(4-\tilde{s})(\tilde{t}^2-\tilde{s}^2)+8\tilde{s}^2\tilde{t}\right)}
{\tilde{s}(\tilde{s}+\tilde{t})^2}\\
%P_{2,9}^{(1)} &= \frac{3}{ \tilde{s} (\tilde{s}+\tilde{t})^2 (\tilde{s}+\tilde{u})^2} (4\tilde{u}^3\tilde{t}^3+ 8\tilde{u}^3\tilde{t}^2\tilde{s} + 4\tilde{u}^3\tilde{t}\tilde{s}^2 + 8\tilde{u}^2\tilde{t}^3\tilde{s} + 15 \tilde{u}^2\tilde{t}^2\tilde{s}^2 \\
%                      &+ 4 \tilde{u}^2\tilde{t}^2\tilde{s} + 8\tilde{u}^2\tilde{t}\tilde{s}^3 + 8\tilde{u}^2\tilde{t}\tilde{s}^2+ \tilde{u}^2\tilde{s}^4 - 4\tilde{u}^2\tilde{s}^3 + 4\tilde{u}\tilde{t}^3\tilde{s}^2 + 8\tilde{u}\tilde{t}^2\tilde{s}^3 + 8 \tilde{u}\tilde{t}^2\tilde{s}^2 \\
%                      &+ 8\tilde{u}\tilde{t}\tilde{s}^4 + 16\tilde{u}\tilde{t}\tilde{s}^3 + 4\tilde{u}\tilde{s}^5 - 8\tilde{u}\tilde{s}^4 + \tilde{t}^2\tilde{s}^4 - 4\tilde{t}^2\tilde{s}^3 + 4\tilde{t}\tilde{s}^5 - 8\tilde{t}\tilde{s}^4 + 3\tilde{s}^6 -12\tilde{s}^5) \\
P_{2,9}^{(1)} &= \frac{3}{ \tilde{s} (\tilde{s}+\tilde{t})^2
(\tilde{s}+\tilde{u})^2} \left\{4\tilde{u}(\tilde{s}+\tilde{t})^2
[2\tilde{s}^2+\tilde{t}(\tilde{s}+\tilde{u})^2] +
8\tilde{s}^4\tilde{t}+4\tilde{s}^2\tilde{t}\tilde{u}
(\tilde{s}^2+2\tilde{u}) \right. \\
& \left. \qquad\qquad\qquad\qquad + \tilde{s} (\tilde{s}-4)
[\tilde{s}^2[(\tilde{s}+\tilde{t})^2+(\tilde{s}+\tilde{u})^2] +
\tilde{s}^3 (2\rho-\tilde{s}) - \tilde{t}^2\tilde{u}^2] \right\} \\
P_{2,10}^{(1)} &= \frac{3 \tilde{s} \tilde{u} (4 - \tilde{s})}{2} \\
P_{2,11}^{(1)} &= \frac{3 \tilde{s} \tilde{t} (4 - \tilde{s})}{2} \\
P_{2,12}^{(1)} &= -\frac{3 \tilde{t} \tilde{u} \left(4\tilde{t}\tilde{u} -\tilde{s}^2 + 12\tilde{s} \right)}{2 \tilde{s}}\, . \\
\end{split}
\end{align}
This result agrees with the one presented in Ref. \cite{Spira:1995es,Spira:1995rr,ptLO1}.

\begin{figure}[h]
\begin{center}
\includegraphics[width=0.8\textwidth]{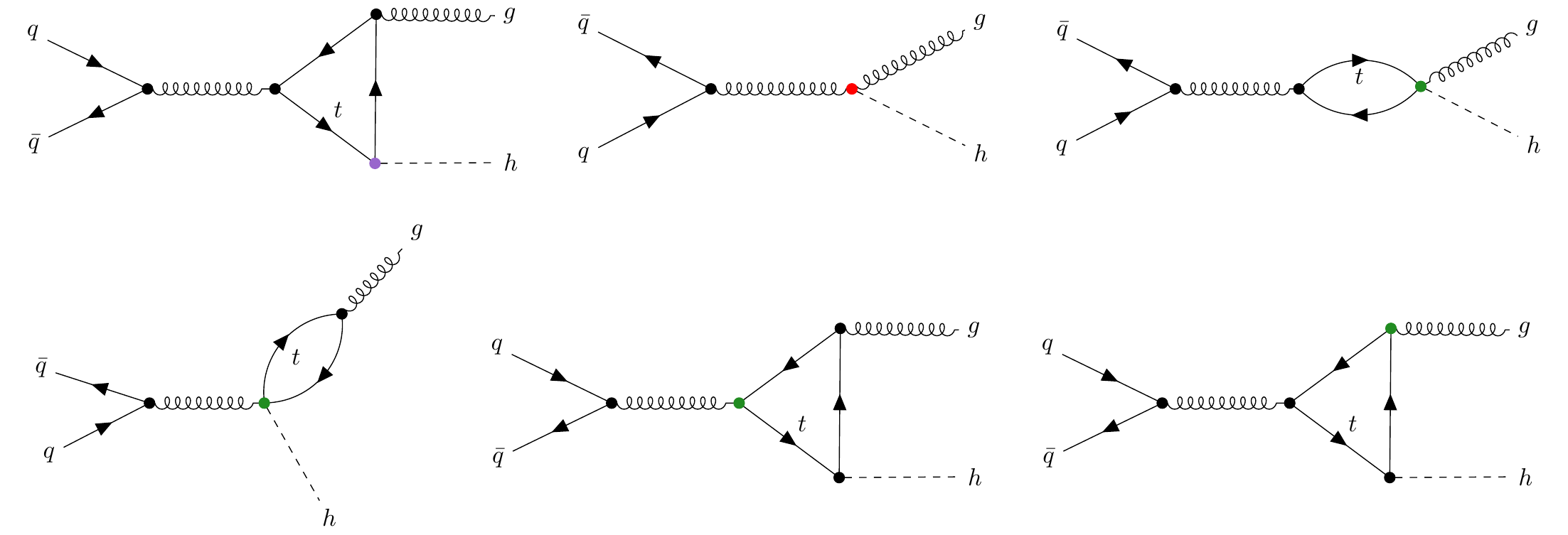}
\end{center}
\caption{\label{fig:qq} Generic diagrams of the $q\bar{q}$-channel
diagrams in the SMEFT.  The color coding is the same as in
Figs.~\ref{fig:SM}--\ref{fig:chromo}.}
\end{figure}

The coefficients for the contribution arising from the chromomagnetic
operator read
\begin{equation}
\begin{array}{llll}
P_{1,1}^{(3)} &= 6 \rho^2 \left(1 - 2 \log\frac{\mu_R^2}{m_t^2}\right) & P_{1,2}^{(3)}  &= -6 \tilde{t} \tilde{u} \\
P_{1,3}^{(3)} &= -6 \tilde{s} \tilde{u} & P_{1,4}^{(3)}  &= -6 \tilde{s} \tilde{t} \\
P_{1,5}^{(3)} &= -6 \rho^2 & P_{1,6}^{(3)}  &= -3 \left(2 \rho - \tilde{t} \tilde{u}\right) \\
P_{1,7}^{(3)} &= -3 \left(2 \rho - \tilde{s} \tilde{u}\right) & P_{1,8}^{(3)}  &= -3 \left(2 \rho - \tilde{s} \tilde{t}\right) \\
P_{1,9}^{(3)} &= 18 \rho & P_{1,10}^{(3)}  &= -3 \tilde{s} \tilde{u} (\rho+\tilde{t}) \\
P_{1,11}^{(3)} &= -3 \tilde{s} \tilde{t} (\rho+ \tilde{u}) & P_{1,12}^{(3)} &= -3 \tilde{t} \tilde{u} (\rho + \tilde{s}) \\
\end{array}
\end{equation}
\begin{align}\label{eq:pctg}
\begin{split}
%P_{2,1}^{(3)}  &= \frac{6 \tilde{s} \left(\tilde{s}^3+\tilde{s}^2 (\tilde{t}+\tilde{u})-\tilde{s} \tilde{t} \tilde{u}-\tilde{t} \tilde{u} (\tilde{t}+\tilde{u})\right)}{(\tilde{s}+\tilde{t}) (\tilde{s}+\tilde{u}) } - 12 \tilde{s}^2 \log\frac{\mu_R^2}{m_t^2} \\
P_{2,1}^{(3)}  &= \frac{6 \rho \tilde{s} \left(\tilde{s}^2 -
\tilde{t}\tilde{u}\right)}{(\tilde{s}+\tilde{t}) (\tilde{s}+\tilde{u}) } - 12 \tilde{s}^2 \log\frac{\mu_R^2}{m_t^2} \\
P_{2,2}^{(3)}  &= 0 \\
%P_{2,3}^{(3)}  &= \frac{6 \tilde{t} \tilde{u} \left(\tilde{s}^2+2 \tilde{s} (\tilde{t}+\tilde{u})+\tilde{u} (\tilde{t}+\tilde{u})\right)}{(\tilde{s}+\tilde{u})^2} \\
P_{2,3}^{(3)}  &= \frac{6 \tilde{t} \tilde{u} \left(\rho^2-\tilde{t}
(\tilde{t}+\tilde{u})\right)}{(\tilde{s}+\tilde{u})^2} \\
%P_{2,4}^{(3)}  &= \frac{6 \tilde{t} \tilde{u} \left(\tilde{s}^2+2 \tilde{s} (\tilde{t}+\tilde{u})+\tilde{t} (\tilde{t}+\tilde{u})\right)}{(\tilde{s}+\tilde{t})^2} \\
P_{2,4}^{(3)}  &= \frac{6 \tilde{t} \tilde{u} \left(\rho^2-\tilde{u}
(\tilde{t}+\tilde{u})\right)}{(\tilde{s}+\tilde{t})^2} \\
%P_{2,5}^{(3)}  &= -\frac{6}{(\tilde{s}+\tilde{t})^2 (\tilde{s}+\tilde{u})^2} (\tilde{s}^6+2 \tilde{s}^5 (\tilde{t}+\tilde{u})+\tilde{s}^4 \left(\tilde{t}^2+6 \tilde{t} \tilde{u}+\tilde{u}^2\right)+8 \tilde{s}^3 \tilde{t} \tilde{u} (\tilde{t}+\tilde{u})\\
%        &+ \tilde{s}^2 \tilde{t} \tilde{u} \left(6 \tilde{t}^2+11 \tilde{t} \tilde{u}+6 \tilde{u}^2\right)+2 \tilde{s} \tilde{t} \tilde{u} (\tilde{t}+\tilde{u})^3+\tilde{t}^2 \tilde{u}^2 (\tilde{t}+\tilde{u})^2)\\
P_{2,5}^{(3)}  &= -\frac{6\rho}{(\tilde{s}+\tilde{t})^2
(\tilde{s}+\tilde{u})^2} (\rho\tilde{s}^2
(\tilde{s}^2+4\tilde{t}\tilde{u})+2\tilde{s}\tilde{t}\tilde{u}
(\rho-\tilde{s})^2-\tilde{t}^2\tilde{u}^2(2\tilde{s}-\rho)) \\
P_{2,6}^{(3)}  &= -6 (2\tilde{s}-\rho)\\
%P_{2,7}^{(3)}  &= -\frac{3 \left(2 \tilde{s} \left(\tilde{s}^3+\tilde{s}^2 (\tilde{u}-\tilde{t})+\tilde{s} \tilde{u} (2 \tilde{t}+\tilde{u})+\tilde{u}^2 (\tilde{t}+\tilde{u})\right)+\tilde{t} \tilde{u} (\tilde{s}+\tilde{u})^2 (\tilde{s}+2 (\tilde{t}+\tilde{u}))\right)}{\tilde{s} (\tilde{s}+\tilde{u})^2}\\
P_{2,7}^{(3)}  &= -\frac{3 \left\{ (\tilde{s}+\tilde{u})^2
\left[2\rho(\tilde{s}+\tilde{t}\tilde{u})-\tilde{s}\tilde{t}\tilde{u}\right]
-4\tilde{s}^2[\tilde{s}(\rho-\tilde{s})+\tilde{u}^2]\right\}}{\tilde{s}
(\tilde{s}+\tilde{u})^2}\\
%P_{2,8}^{(3)}  &= -\frac{3 \left(2 \tilde{s} \left(\tilde{s}^3+\tilde{s}^2 (\tilde{t}-\tilde{u})+\tilde{s} \tilde{t} (\tilde{t}+2 \tilde{u})+\tilde{t}^2 (\tilde{t}+\tilde{u})\right)+\tilde{t} \tilde{u} (\tilde{s}+\tilde{t})^2 (\tilde{s}+2 (\tilde{t}+\tilde{u}))\right)}{\tilde{s} (\tilde{s}+\tilde{t})^2}\\
P_{2,8}^{(3)}  &= -\frac{3 \left\{ (\tilde{s}+\tilde{t})^2
\left[2\rho(\tilde{s}+\tilde{t}\tilde{u})-\tilde{s}\tilde{t}\tilde{u}\right]
-4\tilde{s}^2[\tilde{s}(\rho-\tilde{s})+\tilde{t}^2]\right\}}{\tilde{s}
(\tilde{s}+\tilde{t})^2}\\
%P_{2,9}^{(3)}  &= \frac{3}{ \tilde{s} (\tilde{s}+\tilde{t})^2 (\tilde{s}+\tilde{u})^2} (2 \tilde{s} (3 \tilde{s}^5+3 \tilde{s}^4 (\tilde{t}+\tilde{u})-\tilde{s}^3 \left(\tilde{t}^2-8 \tilde{t} \tilde{u}+\tilde{u}^2\right)\\
%        &-\tilde{s}^2 \left(\tilde{t}^3-5 \tilde{t}^2 \tilde{u}-5 \tilde{t} \tilde{u}^2+\tilde{u}^3\right)+\tilde{s} \tilde{t} \tilde{u} \left(2 \tilde{t}^2+3 \tilde{t} \tilde{u}+2 \tilde{u}^2\right)+\tilde{t}^2 \tilde{u}^2 (\tilde{t}+\tilde{u})) \\
%        &+\tilde{t} \tilde{u} (\tilde{s}+\tilde{t})^2 (\tilde{s}+\tilde{u})^2 (\tilde{s}+2 (\tilde{t}+\tilde{u}))) \\
P_{2,9}^{(3)}  &= \frac{3}{ \tilde{s} (\tilde{s}+\tilde{t})^2
(\tilde{s}+\tilde{u})^2} \left\{2 \tilde{s} \left[3 \rho\tilde{s}^4+
\tilde{s} (2\tilde{t}\tilde{u}-\rho\tilde{s})(\rho-\tilde{s})^2 +
2\tilde{s}^2\tilde{t}\tilde{u}(\tilde{s}+4\rho)
\right. \right. \\
	& \left. \left. \quad\qquad\qquad\qquad\qquad
+\tilde{t}^2\tilde{u}^2 (\rho-2\tilde{s})\right]
+\tilde{t} \tilde{u}
(\tilde{s}+\tilde{t})^2 (\tilde{s}+\tilde{u})^2 (2\rho - \tilde{s})\right\} \\
P_{2,10}^{(3)}  &= -3 \tilde{s} \tilde{u} (\tilde{s}-\tilde{u}) \\
P_{2,11}^{(3)}  &= -3 \tilde{s} \tilde{t} (\tilde{s}-\tilde{t}) \\
%P_{2,12}^{(3)}  &= -\frac{3 \tilde{t} \tilde{u} \left(6 \tilde{s} (\tilde{t}+\tilde{u})+\tilde{t} \tilde{u} (\tilde{s}+2 (\tilde{t}+\tilde{u}))\right)}{2 \tilde{s}} \\
P_{2,12}^{(3)}  &= -\frac{3 \tilde{t} \tilde{u} \left[2 (\rho-\tilde{s})
(3\tilde{s}+\tilde{t}\tilde{u}) + \tilde{s}
\tilde{t}\tilde{u}\right]}{2 \tilde{s}}\, . \\
\end{split}
\end{align}
The $\log \frac{\mu_R^2}{m_t^2} $ terms arise from the absorption of the $1/\epsilon$ divergence in the renormalisation of the $c_g$ coupling.
For the effective Higgs coupling to gluons we directly present the expressions of the $C_1$ and $C_2$ form factors, which read
\begin{align}
\begin{split}
C_1^{(2)}(s,t,u;m_t) &= 12\\
C_2^{(2)}(s,t,u;m_t) & = \frac{12 s^2}{\mh^4},
\end{split}
\end{align}
which correspond also to the HTL of the SM result
multiplied by a factor of 12. For completeness we report the HTL
also for the contribution of the chromomagnetic operator
\begin{equation}
C_i^{(3)}\to \frac{1}{2}(1-\ln \f{\mu_R^2}{m_t^2})\,C_i^{(2)}~~~~~~~~~i=1,2.\nn
\end{equation}
The final results for the form factors $C_i$ read
\begin{equation}
C_i(s,t,u;m_t)=c_1 C^{(1)}_i(s,t,u;m_t) + c_{2}(\mu_R) C^{(2)}_i(s,t,u;m_t) +Re(c_{3})\f{m_t^2}{v^2} C^{(3)}_i(s,t,u;m_t).
\end{equation}
We now move to the $q{\bar q}$ channel:
\begin{equation}
  q(p_1)+{\bar q}(p_2)\to g(p_3)+H(q)\, .\nn
\end{equation}
The contributing generic Feynman diagrams are depicted in Fig.~\ref{fig:qq}.

The corresponding amplitude can be decomposed as
\begin{equation}
  {\cal M}_{q{\bar q}\to gH}(p_1,p_2,p_3) =\frac{ig_S^3}{16 \pi^2 vs}
{\bar v}(p_2)\gamma^\mu T^a\,u(p_1)\left[g_{\mu\nu}(p_1+p_2)\cdot p_3 -
p_{3\mu} (p_1+p_2)_\nu \right]\epsilon^\nu(p_3)\,D(p_1,p_2,p_3)\, .
\end{equation}

We again present the results in the form
\begin{equation}
D^{(\alpha)}(p_1,p_2,p_3)= \sum_{j=1}^5 P^{(\alpha)}_j T_j~~~~~~~~~~\alpha=1,2,3
\end{equation}
with the basis of the scalar integrals:
\begin{equation}
\begin{array}{llll}
T_1 &= 1 & &  \\
T_2 &= 2[1-g(\tau_s)] & T_3 &= 2[1-g(\tau_H)] \\
T_4 &= 2 f(\tau_s) & T_5 &= 2 f(\tau_H)\, . \\
\end{array}
\end{equation}
The coefficients corresponding to the SM and the Yukawa modifying operator in the SMEFT case read:
\begin{align}
\begin{split}
P_{1}^{(1)} &= \frac{8}{(\tilde{s}-\rho)}  \\
P_{2}^{(1)} &=  \frac{8\tilde{s}}{(\tilde{s}-\rho)^2}\\
P_{3}^{(1)} &= -\frac{8\tilde{s}}{(\tilde{s}-\rho)^2} \\
P_{4}^{(1)} &= -\frac{4 \left(4+\tilde{s}-\rho\right)}{(\tilde{s}-\rho)^2} \\
P_{5}^{(1)} &= \frac{4 \left(4+\tilde{s}-\rho\right)}{(\tilde{s}-\rho)^2}\, . \\
\end{split}
\end{align}
The contribution of the chromomagnetic operator reads:
\begin{align}
\begin{split}
P_{1}^{(3)} &= \frac{4 [\rho + 2 (\tilde{s}-\rho) \ln\frac{\mu_R^2}{m_t^2}]}
{(\tilde{s}-\rho)}  \\
P_{2} ^{(3)} &= \frac{2 \left[2 \tilde{s}^2+(\tilde{s}-\rho)^2\right]}
{(\tilde{s}-\rho)^2}\\
P_{3}^{(3)}  &= -\frac{4\rho \left(2\tilde{s}-\rho\right)}
{(\tilde{s}-\rho)^2} \\
P_{4}^{(3)}  &= -\frac{8 (2\tilde{s}-\rho)}{(\tilde{s}-\rho)^2} \\
P_{5}^{(3)}  &= \frac{8 (2\tilde{s}-\rho)}{(\tilde{s}-\rho)^2}.
\end{split}
\end{align}

Again, we finalise the results by presenting the amplitude for the
point-like Higgs coupling to gluons which corresponds to the HTL of the SM:
\begin{equation}
D^{(2)}(p_1,p_2,p_3) = -16\, .
\end{equation}
The HTL of the relevant operators reads:
\begin{equation}
D^{(1)}\to \frac{1}{12}D^{(2)}= -\frac{4}{3}~~~~~~~
D^{(3)}\to \frac{1}{2}(1-\ln \f{\mu_R^2}{m_t^2})D^{(2)} = - 8 \left(1-\ln \f{\mu_R^2}{m_t^2}\right)\, .\nn
\end{equation}
The final expression for the form factor is given by
\begin{equation}
D(p_1,p_2,p_3)=c_1 D^{(1)}(p_1,p_2,p_3) + c_2(\mu_R)  D^{(2)}(p_1,p_2,p_3) +Re(c_3)\f{m_t^2}{v^2} D^{(3)}(p_1,p_2,p_3)\, .
\end{equation}
The result for the $qg$ channel can be obtained by crossing.

The above results allow us to obtain complete predictions for Higgs
boson production at high $p_T$ in the SMEFT. In Ref.~\cite{ours} the
effects of the ${\cal O}_1$ and ${\cal O}_2$ were studied, including the
resummation of the large logarithmic contributions at small $p_T$, but
neglecting the contribution of the chromomagnetic operator.  We thus
focus here on the effect of the chromomagnetic operator at high $p_T$.
We consider $pp$ collisions at $\sqrt{s} = 13$ TeV and use PDF4LHC2015
NLO parton distributions \cite{Butterworth:2015oua,Ball:2014uwa,Dulat:2015mca,Harland-Lang:2014zoa,Gao:2013bia,Carrazza:2015aoa}.
The central value of the renormalization and
factorization scales is fixed to $\mu_F=\mu_R=\sqrt{\mh^2+p_T^2}$.

\begin{figure}[h!]
\begin{center}
\begin{tabular}{cc}
\hspace*{-0.17cm}
\includegraphics[trim = 7mm -7mm 0mm 0mm, width=.36\textheight]{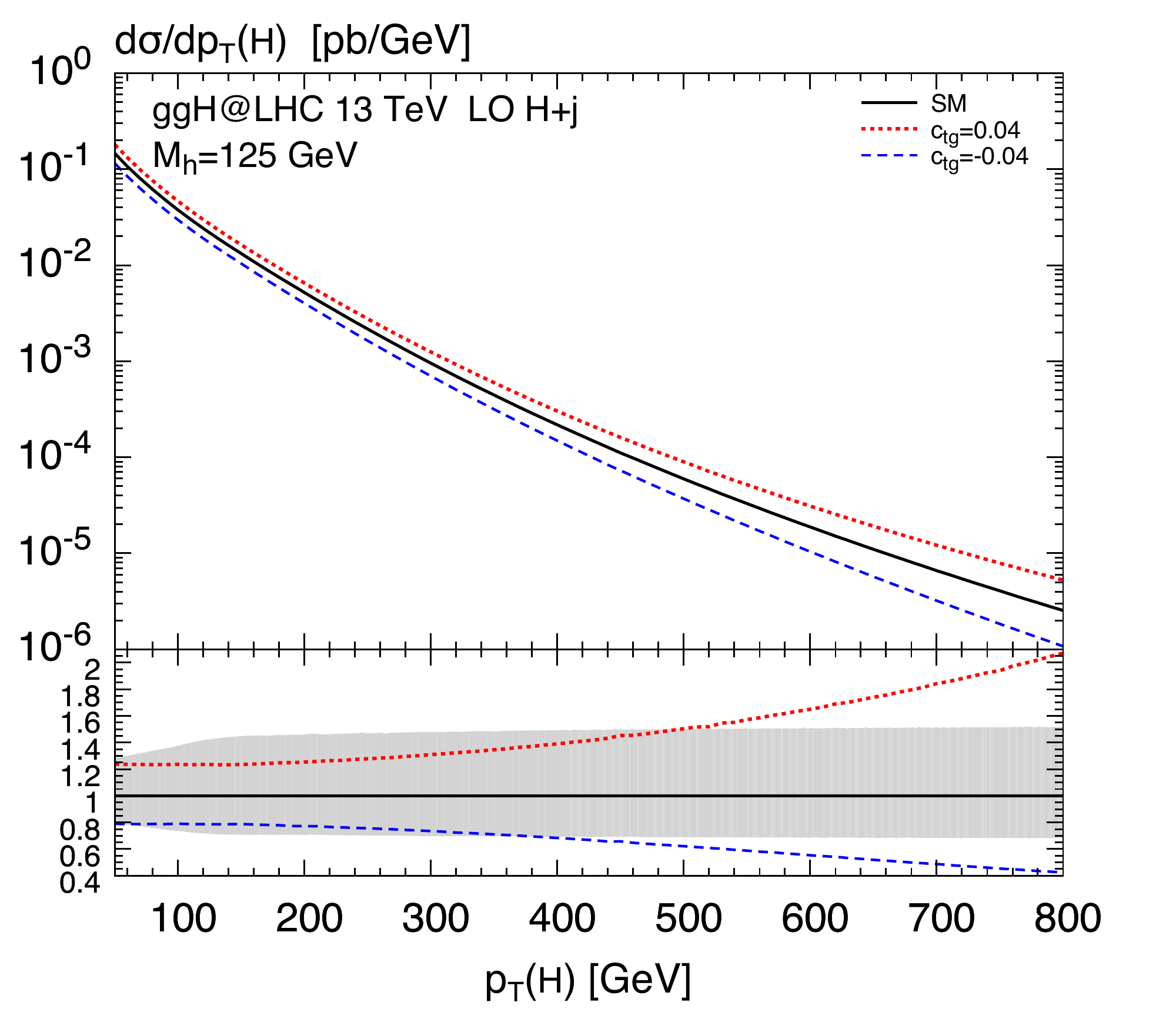}
\end{tabular}
\caption[]{\label{fig:ctg}{Impact of the chromomagnetic operator on the $p_T$ spectrum of the Higgs boson in the region allowed by 
the current experimental constraints.}}
\end{center}
\end{figure}

In Fig.~\ref{fig:ctg} we show the impact of the operator ${\cal O}_3$,
by considering a variation of the coefficient $c_3$ within the range
suggested by the study of Ref.~\cite{Franzosi:2015osa}.  The $p_T$
spectrum including the impact of the chromomagnetic operator is
normalised to the SM result, whose perturbative uncertainty is
estimated with the usual 7-point scale variations.  The numerical
results, obtained with a modified version of the program {\sc Higlu}
\cite{higlu}, show that the chromomagnetic operator can significantly
affect the $p_T$ spectrum, and the effects start to exceed the scale
uncertainty of the SM result around $p_T \approx 400-500 $\,GeV. Our
numerical results agree with those of Ref.~\cite{Maltoni:2016yxb}. A more detailed
study will be presented elsewhere.

\section{Conclusions}
%        ===========
In this work we studied Higgs boson production at large transverse momentum in gluon fusion
within SMEFT.
We provided explicit and compact analytical results for the
one-loop matrix elements of the corresponding partonic processes
$gg,q\bar q\to Hg$ and $gq\to Hq$. The results, which are expressed in terms of standard one-loop scalar integrals,
require the renormalization of the dimension-six point-like Higgs coupling to gluons
in accordance with the related renormalization of the inclusive cross
section \cite{Degrande:2012gr,ours}.  We studied the quantitative impact of
the chromomagnetic operator and found that it can significantly distort the
transverse-momentum spectrum of the Higgs boson at large
$p_T$. Depending on the actual size of the corresponding Wilson
coefficient, this contribution has to be taken into account for a solid
study of relevant dimension-six operators within the SMEFT. Turning this
argument around, the Higgs transverse-momentum distribution will provide
a relevant observable to constrain the Wilson coefficient of the chromomagnetic
operator when significant statistics is accumulated.

\noindent {\bf Acknowledgements.} We would like to thank Fabio Maltoni and Eleni Vryonidou for useful correspondence on the results of Ref.~\cite{Maltoni:2016yxb}.
This research was supported in part by the Swiss National Science Foundation (SNF) under contracts CRSII2-141847,
200020-169041 and by the Research Executive Agency (REA) of the European Union under the
Grant Agreement number PITN--GA--2012--316704 ({\it HiggsTools}).

\appendix

\section{Scalar Integrals}\label{sec:scalint}
%        ================
In this appendix we present the definitions of the functions $f,g,J$
used for our analytical results and their relation to the scalar
one-loop integrals.

\begin{equation}
\label{eq:goftau}
g(\tau) = \begin{cases}
\displaystyle \sqrt{\tau-1}\arcsin \frac{1}{\sqrt{\tau}}  & \tau \ge 1 \\
\displaystyle \frac{\sqrt{1-\tau}}{2}\left[ \ln \frac{1+\sqrt{1-\tau}}
{1-\sqrt{1-\tau}} - i\pi \right]  & \tau < 1
\end{cases} \;.
\end{equation}

\begin{equation}
\label{eq:foftau}
f(\tau) = \begin{cases}
\displaystyle \arcsin^2 \frac{1}{\sqrt{\tau}}  & \tau \ge 1 \\
\displaystyle - \frac{1}{4} \left[ \ln \frac{1+\sqrt{1-\tau}}
{1-\sqrt{1-\tau}} - i\pi \right]^2  & \tau < 1
\end{cases} \;
\end{equation}

\begin{align}\begin{split}
J(\tilde{s},\tilde{t},\tilde{u}) &= I_3(\tilde{s},\tilde{t},\tilde{u},\tilde{s}) + I_3(\tilde{s},\tilde{t},\tilde{u},\tilde{u}) - I_3(\tilde{s},\tilde{t},\tilde{u}, \rho)\\
I_3(\tilde{s},\tilde{t},\tilde{u},\tilde{x}) &= \frac{1}{\tilde{s} \tilde{u}} \frac{2}{\beta_+ - \beta_-} \bigg( \text{Li}_2\left(\frac{\beta_-}{\beta_- - \alpha_-}\right) - \text{Li}_2\left(\frac{\beta_+}{\beta_+ - \alpha_+}\right) + \text{Li}_2\left(\frac{\beta_-}{\beta_- - \alpha_+}\right)\\
& - \text{Li}_2\left(\frac{\beta_+}{\beta_+ - \alpha_-}\right) + \log \left(-\frac{\beta_+}{\beta_-}\right)\log \left(1+ \frac{\tilde{x} \tilde{t}}{\tilde{s} \tilde{u}}\right) \bigg) ,
\end{split}\end{align}
with $\alpha_{\pm} = \frac{1}{2}(1 \pm \sqrt{1 - \frac{4}{\tilde{x}}})$
and $\beta_{\pm} = \frac{1}{2}(1 \pm \sqrt{1 -
\frac{4\tilde{t}}{\tilde{s} \tilde{u}}})$. The function $\text{Li}_2$
denotes the Spence function. The functions $f,g,J$ are related
to the corresponding scalar one-loop integrals as ($n=4-2\epsilon$)
\begin{eqnarray}
B_0(p;m_t,m_t) \!\!\! & = & \!\!\! \int \frac{d^nk}{(2\pi)^n}
\frac{\mu^{2\epsilon}}{(k^2-m_t^2) [(k+p)^2-m_t^2]} \nonumber \\
& = & i \frac{\Gamma(1+\epsilon)}{(4\pi)^2}
\left(\frac{4\pi\mu^2}{m_t^2}\right)^\epsilon \left[ \frac{1}{\epsilon}
+ 2 - 2g(\tau)\right] + {\cal O}(\epsilon)
\quad\qquad (\tau=4 m_t^2/p^2) \nonumber \\
C_0(p_1,p_2;m_t,m_t,m_t) \!\!\! & = & \!\!\! \int \frac{d^nk}{(2\pi)^n}
\frac{\mu^{2\epsilon}}{(k^2-m_t^2) [(k-p_1)^2-m_t^2] [(k+p_2)^2-m_t^2]}
\nonumber \\
& = & \frac{i}{(4\pi)^2 m_t^2}
\left[ -\frac{\tau}{2} f(\tau)\right] + {\cal O}(\epsilon)
\quad\qquad\qquad\qquad\qquad (\tau=4 m_t^2/(2 p_1p_2)) \nonumber \\
D_0(p_1,p_2,p_3;m_t,m_t,m_t,m_t) \!\!\!\!\! & = & \!\!\!\!\!\!\!
\int \frac{d^nk}{(2\pi)^n}
\frac{\mu^{2\epsilon}}{(k^2-m_t^2) [(k+p_1)^2-m_t^2] [(k+p_{12})^2-m_t^2]
[(k+p_{123})^2-m_t^2]} \nonumber \\
& = & \frac{i}{(4\pi)^2 m_t^4}
J\left(\frac{s}{m_t^2},\frac{t}{m_t^2},\frac{u}{m_t^2}\right) + {\cal
O}(\epsilon)
\end{eqnarray}
where $p_{12} = p_1+p_2, p_{123} = p_1+p_2+p_3$ and $s=2 p_1p_2, t =
2p_1p_3, u=2p_2p_3$ where all $p_i~(i=1,2,3)$ correspond to incoming
massless external particles ($p_i^2=0$).

\bibliographystyle{UTPstyle}
\bibliography{biblio}

\providecommand{\href}[2]{#2}\begingroup\raggedright\begin{thebibliography}{10}

\bibitem{ATLASdisc}
{\bf ATLAS} Collaboration, G.~Aad et~al., {\it {Observation of a new particle
  in the search for the Standard Model Higgs boson with the ATLAS detector at
  the LHC}},  {\em Phys. Lett.} {\bf B716} (2013) 1--29,
  [\href{http://xxx.lanl.gov/abs/1207.7214}{{\tt arXiv:1207.7214}}].

\bibitem{CMSdisc}
{\bf CMS} Collaboration, S.~Chatrchyan et~al., {\it {Observation of a new boson
  at a mass of 125 GeV with the CMS experiment at the LHC}},  {\em Phys. Lett.}
  {\bf B716} (2013) 30--61, [\href{http://xxx.lanl.gov/abs/1207.7235}{{\tt
  arXiv:1207.7235}}].

\bibitem{Khachatryan:2016vau}
{\bf ATLAS, CMS} Collaboration, G.~Aad et~al., {\it {Measurements of the Higgs
  boson production and decay rates and constraints on its couplings from a
  combined ATLAS and CMS analysis of the LHC pp collision data at $ \sqrt{s}=7
  $ and 8 TeV}},  {\em JHEP} {\bf 08} (2016) 045,
  [\href{http://xxx.lanl.gov/abs/1606.02266}{{\tt arXiv:1606.02266}}].

\bibitem{Heinemeyer:2013tqa}
{\bf LHC Higgs Cross Section Working Group} Collaboration, J.~R. Andersen
  et~al., {\it {Handbook of LHC Higgs Cross Sections: 3. Higgs Properties}},
  \href{http://xxx.lanl.gov/abs/1307.1347}{{\tt arXiv:1307.1347}}.

\bibitem{YR4}
{\bf LHC Higgs Cross Section Working Group} Collaboration, D.~de~Florian
  et~al., {\it {Handbook of LHC Higgs Cross Sections: 4. Deciphering the Nature
  of the Higgs Sector}},  \href{http://xxx.lanl.gov/abs/1610.07922}{{\tt
  arXiv:1610.07922}}.

\bibitem{HTBook}
{\bf HiggsTools Working Group} Collaboration, M.~Boggia et~al., {\it {The
  HiggsTools Handbook: Concepts and observables for deciphering the Nature of
  the Higgs Sector}},  \href{http://xxx.lanl.gov/abs/1711.09875}{{\tt
  arXiv:1711.09875}}.

\bibitem{dim61}
W.~Buchmuller and D.~Wyler, {\it {Effective Lagrangian Analysis of New
  Interactions and Flavor Conservation}},  {\em Nucl. Phys.} {\bf B268} (1986)
  621--653.

\bibitem{dim62}
B.~Grzadkowski, M.~Iskrzynski, M.~Misiak, and J.~Rosiek, {\it {Dimension-Six
  Terms in the Standard Model Lagrangian}},  {\em JHEP} {\bf 10} (2010) 085,
  [\href{http://xxx.lanl.gov/abs/1008.4884}{{\tt arXiv:1008.4884}}].

\bibitem{atlas1}
{\bf ATLAS} Collaboration, G.~Aad et~al., {\it {Measurements of fiducial and
  differential cross sections for Higgs boson production in the diphoton decay
  channel at $\sqrt{s}=8$ TeV with ATLAS}},  {\em JHEP} {\bf 09} (2014) 112,
  [\href{http://xxx.lanl.gov/abs/1407.4222}{{\tt arXiv:1407.4222}}].

\bibitem{atlas2}
{\bf ATLAS} Collaboration, G.~Aad et~al., {\it {Fiducial and differential cross
  sections of Higgs boson production measured in the four-lepton decay channel
  in $pp$ collisions at $\sqrt{s}$=8 TeV with the ATLAS detector}},  {\em Phys.
  Lett.} {\bf B738} (2014) 234--253,
  [\href{http://xxx.lanl.gov/abs/1408.3226}{{\tt arXiv:1408.3226}}].

\bibitem{Aad:2015lha}
{\bf ATLAS} Collaboration, G.~Aad et~al., {\it {Measurements of the Total and
  Differential Higgs Boson Production Cross Sections Combining the $H\to \gamma
  \gamma$ and $H\to ZZ^* \to 4l$ Decay Channels at $\sqrt{s}$=8??TeV with the
  ATLAS Detector}},  {\em Phys. Rev. Lett.} {\bf 115} (2015), no.~9 091801,
  [\href{http://xxx.lanl.gov/abs/1504.05833}{{\tt arXiv:1504.05833}}].

\bibitem{Aad:2016lvc}
{\bf ATLAS} Collaboration, G.~Aad et~al., {\it {Measurement of fiducial
  differential cross sections of gluon-fusion production of Higgs bosons
  decaying to WW$^{*}\to e\nu\mu\nu$ with the ATLAS detector at $ \sqrt{s}=8 $
  TeV}},  {\em JHEP} {\bf 08} (2016) 104,
  [\href{http://xxx.lanl.gov/abs/1604.02997}{{\tt arXiv:1604.02997}}].

\bibitem{CMSpt}
{\bf CMS} Collaboration, S.~Chatrchyan et~al., {\it {Measurement of
  differential cross sections for Higgs boson production in the diphoton decay
  channel in pp collisions at $\sqrt(s)=8$ TeV}},  {\em arXiv:1508.07819}
  (2015) [\href{http://xxx.lanl.gov/abs/1508.07819}{{\tt arXiv:1508.07819}}].

\bibitem{Khachatryan:2015yvw}
{\bf CMS} Collaboration, V.~Khachatryan et~al., {\it {Measurement of
  differential and integrated fiducial cross sections for Higgs boson
  production in the four-lepton decay channel in pp collisions at $ \sqrt{s}=7
  $ and 8 TeV}},  {\em JHEP} {\bf 04} (2016) 005,
  [\href{http://xxx.lanl.gov/abs/1512.08377}{{\tt arXiv:1512.08377}}].

\bibitem{Khachatryan:2016vnn}
{\bf CMS} Collaboration, V.~Khachatryan et~al., {\it {Measurement of the
  transverse momentum spectrum of the Higgs boson produced in pp collisions at
  $ \sqrt{s}=8 $ TeV using $H \to WW$ decays}},  {\em JHEP} {\bf 03} (2017)
  032, [\href{http://xxx.lanl.gov/abs/1606.01522}{{\tt arXiv:1606.01522}}].

\bibitem{Aaboud:2017oem}
{\bf ATLAS} Collaboration, M.~Aaboud et~al., {\it {Measurement of inclusive and
  differential cross sections in the $H \rightarrow ZZ^* \rightarrow 4\ell$
  decay channel in $pp$ collisions at $\sqrt{s}=13$ TeV with the ATLAS
  detector}},  {\em JHEP} {\bf 10} (2017) 132,
  [\href{http://xxx.lanl.gov/abs/1708.02810}{{\tt arXiv:1708.02810}}].

\bibitem{Aaboud:2018xdt}
{\bf ATLAS} Collaboration, M.~Aaboud et~al., {\it {Measurements of Higgs boson
  properties in the diphoton decay channel with 36 fb$^{-1}$ of $pp$ collision
  data at $\sqrt{s} = 13$ TeV with the ATLAS detector}},
  \href{http://xxx.lanl.gov/abs/1802.04146}{{\tt arXiv:1802.04146}}.

\bibitem{Sirunyan:2017exp}
{\bf CMS} Collaboration, A.~M. Sirunyan et~al., {\it {Measurements of
  properties of the Higgs boson decaying into the four-lepton final state in pp
  collisions at $ \sqrt{s}=13 $ TeV}},  {\em JHEP} {\bf 11} (2017) 047,
  [\href{http://xxx.lanl.gov/abs/1706.09936}{{\tt arXiv:1706.09936}}].

\bibitem{Banfi:2013yoa}
A.~Banfi, A.~Martin, and V.~Sanz, {\it {Probing top-partners in Higgs+jets}},
  {\em JHEP} {\bf 08} (2014) 053,
  [\href{http://xxx.lanl.gov/abs/1308.4771}{{\tt arXiv:1308.4771}}].

\bibitem{Azatov:2016xik}
A.~Azatov, C.~Grojean, A.~Paul, and E.~Salvioni, {\it {Resolving gluon fusion
  loops at current and future hadron colliders}},  {\em JHEP} {\bf 09} (2016)
  123, [\href{http://xxx.lanl.gov/abs/1608.00977}{{\tt arXiv:1608.00977}}].

\bibitem{Brooijmans:2016vro}
G.~Brooijmans et~al., {\it {Les Houches 2015: Physics at TeV colliders - new
  physics working group report}},
  \href{http://xxx.lanl.gov/abs/1605.02684}{{\tt arXiv:1605.02684}}.

\bibitem{Harlander:2016hcx}
R.~V. Harlander, S.~Liebler, and H.~Mantler, {\it {SusHi Bento: Beyond NNLO and
  the heavy-top limit}},  {\em Comput. Phys. Commun.} {\bf 212} (2017)
  239--257, [\href{http://xxx.lanl.gov/abs/1605.03190}{{\tt
  arXiv:1605.03190}}].

\bibitem{Anastasiou:2016hlm}
C.~Anastasiou, C.~Duhr, F.~Dulat, E.~Furlan, T.~Gehrmann, F.~Herzog,
  A.~Lazopoulos, and B.~Mistlberger, {\it {CP-even scalar boson production via
  gluon fusion at the LHC}},  {\em JHEP} {\bf 09} (2016) 037,
  [\href{http://xxx.lanl.gov/abs/1605.05761}{{\tt arXiv:1605.05761}}].

\bibitem{ptdim61}
C.~Grojean, E.~Salvioni, M.~Schlaffer, and A.~Weiler, {\it {Very boosted Higgs
  in gluon fusion}},  {\em JHEP} {\bf 05} (2014) 022,
  [\href{http://xxx.lanl.gov/abs/1312.3317}{{\tt arXiv:1312.3317}}].

\bibitem{ptdim62}
A.~Azatov and A.~Paul, {\it {Probing Higgs couplings with high $p_T$ Higgs
  production}},  {\em JHEP} {\bf 01} (2014) 014,
  [\href{http://xxx.lanl.gov/abs/1309.5273}{{\tt arXiv:1309.5273}}].

\bibitem{ptdim63}
U.~Langenegger, M.~Spira, and I.~Strebel, {\it {Testing the Higgs Boson
  Coupling to Gluons}},  {\em arXiv:1507.01373} (2015)
  [\href{http://xxx.lanl.gov/abs/1507.01373}{{\tt arXiv:1507.01373}}].

\bibitem{ptdim81}
R.~V. Harlander and T.~Neumann, {\it {Probing the nature of the Higgs-gluon
  coupling}},  {\em Phys. Rev.} {\bf D88} (2013) 074015,
  [\href{http://xxx.lanl.gov/abs/1308.2225}{{\tt arXiv:1308.2225}}].

\bibitem{ptdim82}
S.~Dawson, I.~M. Lewis, and M.~Zeng, {\it {Effective field theory for Higgs
  boson plus jet production}},  {\em Phys. Rev.} {\bf D90} (2014), no.~9
  093007, [\href{http://xxx.lanl.gov/abs/1409.6299}{{\tt arXiv:1409.6299}}].

\bibitem{Choudhury:2012np}
D.~Choudhury and P.~Saha, {\it {Higgs production as a probe of anomalous top
  couplings}},  {\em JHEP} {\bf 08} (2012) 144,
  [\href{http://xxx.lanl.gov/abs/1201.4130}{{\tt arXiv:1201.4130}}].

\bibitem{Degrande:2012gr}
C.~Degrande, J.~M. Gerard, C.~Grojean, F.~Maltoni, and G.~Servant, {\it
  {Probing Top-Higgs Non-Standard Interactions at the LHC}},  {\em JHEP} {\bf
  07} (2012) 036, [\href{http://xxx.lanl.gov/abs/1205.1065}{{\tt
  arXiv:1205.1065}}]. [Erratum: JHEP03,032(2013)].

\bibitem{Zhang:2016snc}
C.~Zhang, {\it {Automating Predictions for Standard Model Effective Field
  Theory in MadGraph5 aMC@NLO}},  {\em PoS} {\bf RADCOR2015} (2016) 101,
  [\href{http://xxx.lanl.gov/abs/1601.03994}{{\tt arXiv:1601.03994}}].

\bibitem{Maltoni:2016yxb}
F.~Maltoni, E.~Vryonidou, and C.~Zhang, {\it {Higgs production in association
  with a top-antitop pair in the Standard Model Effective Field Theory at NLO
  in QCD}},  {\em JHEP} {\bf 10} (2016) 123,
  [\href{http://xxx.lanl.gov/abs/1607.05330}{{\tt arXiv:1607.05330}}].

\bibitem{Deutschmann:2017qum}
N.~Deutschmann, C.~Duhr, F.~Maltoni, and E.~Vryonidou, {\it {Gluon-fusion Higgs
  production in the Standard Model Effective Field Theory}},  {\em JHEP} {\bf
  12} (2017) 063, [\href{http://xxx.lanl.gov/abs/1708.00460}{{\tt
  arXiv:1708.00460}}].

\bibitem{ours}
M.~Grazzini, A.~Ilnicka, M.~Spira, and M.~Wiesemann, {\it {Modeling BSM effects
  on the Higgs transverse-momentum spectrum in an EFT approach}},  {\em JHEP}
  {\bf 03} (2017) 115, [\href{http://xxx.lanl.gov/abs/1612.00283}{{\tt
  arXiv:1612.00283}}].

\bibitem{SILH0}
G.~F. Giudice, C.~Grojean, A.~Pomarol, and R.~Rattazzi, {\it {The
  Strongly-Interacting Light Higgs}},  {\em JHEP} {\bf 06} (2007) 045,
  [\href{http://xxx.lanl.gov/abs/hep-ph/0703164}{{\tt hep-ph/0703164}}].

\bibitem{SILH}
R.~Contino, M.~Ghezzi, C.~Grojean, M.~Muhlleitner, and M.~Spira, {\it
  {Effective Lagrangian for a light Higgs-like scalar}},  {\em JHEP} {\bf 07}
  (2013) 035, [\href{http://xxx.lanl.gov/abs/1303.3876}{{\tt
  arXiv:1303.3876}}].

\bibitem{our_Moriond}
M.~Grazzini, A.~Ilnicka, M.~Spira, and M.~Wiesemann, {\it {Effective Field
  Theory for Higgs properties parametrisation: the transverse momentum spectrum
  case}},  {\em {52nd Rencontres de Moriond QCD 2017, La Thuile}} (2017)
  [\href{http://xxx.lanl.gov/abs/1705.05143}{{\tt arXiv:1705.05143}}].

\bibitem{our_discrete}
M.~Grazzini, A.~Ilnicka, M.~Spira, and M.~Wiesemann, {\it {Effective Field
  Theory in quest to parametrise Higgs properties: the transverse momentum
  spectrum case}},  {\em J. Phys. Conf. Ser.} {\bf 873} (2017), no.~1 012050.

\bibitem{Spira:1995es}
M.~Spira, {\it {Radiative QCD corrections to decay and production of Higgs
  bosons at e+e- and pp accelerators (in German)}}, , PhD thesis, Aachen, Tech.
  Hochsch.,
\newblock 1993.

\bibitem{Spira:1995rr}
M.~Spira, A.~Djouadi, D.~Graudenz, and P.~M. Zerwas, {\it {Higgs boson
  production at the LHC}},  {\em Nucl. Phys.} {\bf B453} (1995) 17--82,
  [\href{http://xxx.lanl.gov/abs/hep-ph/9504378}{{\tt hep-ph/9504378}}].

\bibitem{ptLO1}
R.~K. Ellis, I.~Hinchliffe, M.~Soldate, and J.~J. van~der Bij, {\it {Higgs
  Decay to tau+ tau-: A Possible Signature of Intermediate Mass Higgs Bosons at
  the SSC}},  {\em Nucl. Phys.} {\bf B297} (1988) 221.

\bibitem{Butterworth:2015oua}
J.~Butterworth et~al., {\it {PDF4LHC recommendations for LHC Run II}},  {\em J.
  Phys.} {\bf G43} (2016) 023001,
  [\href{http://xxx.lanl.gov/abs/1510.03865}{{\tt arXiv:1510.03865}}].

\bibitem{Ball:2014uwa}
{\bf NNPDF} Collaboration, R.~D. Ball et~al., {\it {Parton distributions for
  the LHC Run II}},  {\em JHEP} {\bf 04} (2015) 040,
  [\href{http://xxx.lanl.gov/abs/1410.8849}{{\tt arXiv:1410.8849}}].

\bibitem{Dulat:2015mca}
S.~Dulat, T.-J. Hou, J.~Gao, M.~Guzzi, J.~Huston, P.~Nadolsky, J.~Pumplin,
  C.~Schmidt, D.~Stump, and C.~P. Yuan, {\it {New parton distribution functions
  from a global analysis of quantum chromodynamics}},  {\em Phys. Rev.} {\bf
  D93} (2016), no.~3 033006, [\href{http://xxx.lanl.gov/abs/1506.07443}{{\tt
  arXiv:1506.07443}}].

\bibitem{Harland-Lang:2014zoa}
L.~A. Harland-Lang, A.~D. Martin, P.~Motylinski, and R.~S. Thorne, {\it {Parton
  distributions in the LHC era: MMHT 2014 PDFs}},  {\em Eur. Phys. J.} {\bf
  C75} (2015), no.~5 204, [\href{http://xxx.lanl.gov/abs/1412.3989}{{\tt
  arXiv:1412.3989}}].

\bibitem{Gao:2013bia}
J.~Gao and P.~Nadolsky, {\it {A meta-analysis of parton distribution
  functions}},  {\em JHEP} {\bf 07} (2014) 035,
  [\href{http://xxx.lanl.gov/abs/1401.0013}{{\tt arXiv:1401.0013}}].

\bibitem{Carrazza:2015aoa}
S.~Carrazza, S.~Forte, Z.~Kassabov, J.~I. Latorre, and J.~Rojo, {\it {An
  Unbiased Hessian Representation for Monte Carlo PDFs}},  {\em Eur. Phys. J.}
  {\bf C75} (2015), no.~8 369, [\href{http://xxx.lanl.gov/abs/1505.06736}{{\tt
  arXiv:1505.06736}}].

\bibitem{Franzosi:2015osa}
D.~Buarque~Franzosi and C.~Zhang, {\it {Probing the top-quark chromomagnetic
  dipole moment at next-to-leading order in QCD}},  {\em Phys. Rev.} {\bf D91}
  (2015), no.~11 114010, [\href{http://xxx.lanl.gov/abs/1503.08841}{{\tt
  arXiv:1503.08841}}].

\bibitem{higlu}
M.~Spira, {\it {HIGLU: A program for the calculation of the total Higgs
  production cross-section at hadron colliders via gluon fusion including QCD
  corrections}},  {\em arXiv:hep-ph/9510347} (1995)
  [\href{http://xxx.lanl.gov/abs/hep-ph/9510347}{{\tt hep-ph/9510347}}].

\end{thebibliography}\endgroup

\end{document}